\documentclass[11pt]{article}
\usepackage{amsmath,amssymb,color}

\usepackage{amsfonts}

\newcommand{\hoch}[1]{$\, ^{#1}$}

\newcommand{\be}{\begin{equation}}
\newcommand{\ee}{\end{equation}}
\newcommand{\bea}{\setlength\arraycolsep{2pt} \begin{eqnarray}}
\newcommand{\eea}{\end{eqnarray}}
\newcommand{\nn}{\nonumber}

\def\ft#1#2{{\textstyle{\frac{\scriptstyle #1}{\scriptstyle #2} } }}
\def\fft#1#2{{\frac{#1}{#2}}}

\def\0{{\sst{(0)}}}
\def\1{{\sst{(1)}}}
\def\2{{\sst{(2)}}}
\def\3{{\sst{(3)}}}
\def\4{{\sst{(4)}}}
\def\5{{\sst{(5)}}}
\def\6{{\sst{(6)}}}
\def\7{{\sst{(7)}}}
\def\8{{\sst{(8)}}}
\def\sst#1{{\scriptscriptstyle #1}}

\def\del{{\partial}}

\def\cG{{{\cal G}}}

\begin{document}

\hfill{CAS-KITPC/ITP-255}

\vspace{25pt}
\begin{center}
{\large {\bf On Black Hole Stability in Critical Gravities}}

\vspace{15pt}

Haishan Liu\hoch{1,2}, H. L\"u\hoch{3,4} and Mingxing Luo\hoch{1}

\vspace{10pt}

\hoch{1}{\it Zhejiang Institute of Modern Physics, Department of
Physics\\
Zhejiang University, Hangzhou, 310027}

\vspace{10pt}

\hoch{2}{\it Kavli Institute for Theoretical Physics China, CAS,
Beijing 100190, China}

\vspace{10pt}

\hoch{3}{\it China Economics and Management Academy\\
Central University of Finance and Economics, Beijing 100081}

\vspace{10pt}

\hoch{4}{\it Institute for Advanced Study, Shenzhen University,
Nanhai Ave 3688, Shenzhen 518060}

\vspace{40pt}

\underline{ABSTRACT}
\end{center}

We consider extended cosmological gravities with Ricci tensor and
scalar squared terms in diverse dimensions.  These theories admit
solutions of Einstein metrics, including the
Schwarzschild-Tangherlini AdS black holes, whose mass and entropy
vanish at the critical point. We perform linearized analysis around
the black holes and show that in general the spectrum consists of
the usual spin-2 massless and ghost massive modes.  We demonstrate
that there is no exponentially-growing tachyon mode in the black
holes.  At the critical point, the massless spin-2 modes have zero
energy whilst the massive spin-2 modes are replaced by the log
modes. There always exist certain linear combination of massless and
log modes that has negative energy. Thus the stability of the black
holes requires that the log modes to be truncated out by the
boundary condition.

\vspace{15pt}

\thispagestyle{empty}





\newpage

Black hole physics is an important research topic in General
Relativity.  Here an important question is the classical stability
of black holes.  The simplest black hole is the static and
spherically symmetric one, which was first obtained by
Schwarzschild. The classical stability of the Schwarzschild black
hole was established long ago \cite{old1,old2,old3}. The stability
condition for static black holes in Einstein gravity with a
cosmological constant in higher dimensions was studied in
\cite{gibhar}. (See also \cite{ik1,ik2}.) In higher dimensions,
there exist inhomogeneous Einstein metrics on manifolds that are
topologically spheres \cite{bohm}.  The stability of black holes
with these level surfaces was discussed in detail in \cite{gihapo}.

It is natural to investigate the stability of black holes in
extended gravities with higher-derivative curvature terms.  One
difficulty associated with this line of research is, when
polynomials of the Riemann tensor are added to the Lagrangian, it is
difficult to construct analytical black hole solutions, even for the
static cohomogeneity one configuration. When the coupling of the
curvature squared terms are small, perturbative construction were
discussed in \cite{noo,no}. Furthermore, the introduction of
higher-derivative terms gives rise to ghost-like massive graviton in
general \cite{kstelle}. This automatically implies instability of
the black hole, even if a solution exists. A counterexample is when
higher-derivative terms form a topological Gauss-Bonnet combination.
In such case, both problems appear to be resolved. The theory is
ghost-free, from the linearized analysis of the vacuum. It can be
shown that the propagator of the graviton is of two derivatives
rather than four derivatives. Actually, there exists explicit
analytical black hole solution in Gauss-Bonnet gravity
\cite{boudes,cai}. However it was shown in \cite{bdg} that these
black holes with small mass are not stable. Moreover, the
Gauss-Bonnet term vanishes at $D=3$ and is a total derivative in
$D=4$. It is only relevant for $D\ge5$.

Recently, critical gravities in four \cite{lpcritical} and higher
dimensions \cite{dllpst} were obtained, generalizing the results in
$D=3$ \cite{strom,bhrt1}. These theories are extended Einstein
gravities with cosmological constants and curvature square terms.
These theories admit AdS spacetime as vacuum solutions. Linearized
analysis around the vacua shows that for an appropriate choice of
parameters, the ghost-like massive graviton is absent. It is
replaced by log modes of a fourth-order operator that is the square
of the second-order operator of the massless spin-2 graviton. These
modes satisfy different asymptotic behavior from the usual massless
and massive modes, and they can be consistently truncated out by
imposing the standard boundary condition. The explicit forms of log
modes were constructed recently \cite{af,ggst,bhrt}. Although an
explicit calculation shows that the log modes in four dimensions may
have positive energy \cite{lpcritical}, it was pointed that there
are negative norm states associated with certain linear combination
of the massless and log modes \cite{porrati}. Thus ghost-free
condition requires the truncation of log modes by the boundary
condition. The purpose of this paper is to generalize the vacuum
analysis to static black holes. We find that the critical point
derived from black hole backgrounds is the same as that from the
vacua. We show that black hole solutions do not suffer from the
tachyon instability, but the subtle instability associated with log
modes persists, which should hence be truncated out.

Let us consider the action in $n\ge 3$ dimensions
\begin{equation}
I = \fft{1}{2 \kappa^{2}} \int \sqrt{-g} d^{n}x [R - (n-2)\Lambda_0
+\alpha R^2 +  \beta R^{\mu\nu} R_{\mu\nu}]\,.\label{genaction}
\end{equation}
Note that the Gauss-Bonnet term is not included, as it is not
essential for criticality. In fact, there is no critical phenomenon
with purely the Gauss-Bonnet term, since the corresponding
propagators of linearized modes are of the second order.  In
addition, as we remarked earlier, the Gauss-Bonnet term vanishes at
$D=3$ and it is a total derivative in $D=4$. Furthermore, there is
no known analytical black hole solution when we include both the
Gauss-Bonnet and other curvature square terms. As we shall see
presently, by excluding the Gauss-Bonnet term, these theories admit
solutions of Einstein metrics, which include the
Schwarzschild-Tangherlini AdS black holes.

    The equations of motion are given by
\begin{equation}
\cG_{\mu\nu} + \alpha E_{\mu\nu}^\1 + \beta
E_{\mu\nu}^\2=0\,,\label{eom}
\end{equation}
where
\begin{eqnarray}
\cG_{\mu\nu} &=& R_{\mu\nu} -\ft12 R\,g_{\mu\nu} +
\ft12(n-2)\Lambda_0\, g_{\mu\nu}\,,\label{cGdef}\\
E_{\mu\nu}^\1 &=& 2R(R_{\mu\nu} -\ft14 R\, g_{\mu\nu}) +
2g_{\mu\nu}\, \square R -2\nabla_\mu\nabla_\nu R\,,\label{emunu2}\\
E_{\mu\nu}^\2 &=& -\ft12 R^{\rho\sigma} R_{\rho\sigma}\, g_{\mu\nu}
+ 2 R_{\mu\rho\nu\sigma} R^{\rho\sigma}+\Box R_{\mu\nu} + \ft12\Box
R\,g_{\mu\nu} - \nabla_\mu \nabla_{\nu} R\,.\label{emunu1}
\end{eqnarray}
The trace part of (\ref{eom}) is given by
\begin{eqnarray}
&& n(n-2)\Lambda_{0} -  (n-2) R -\alpha \Big((n-4) R^2 - 4(n-1) \Box
R\Big) \cr &&- \beta \Big( (n-4) R_{\mu\nu} R^{\mu\nu} - n \Box
R\Big)
 =0\,.\label{eomtrace}
\end{eqnarray}

Let us consider the Einstein metrics, satisfying
\begin{equation}
R_{\mu\nu} = \Lambda g_{\mu\nu}\,,\qquad R=n
\Lambda\,.\label{einstein}
\end{equation}
The corresponding Einstein tensor is given by
\begin{equation}
\widetilde \cG_{\mu\nu} = R_{\mu\nu} - \ft12 R\, g_{\mu\nu} + \ft12
(n-2) \Lambda\, g_{\mu\nu}\,.
\end{equation}
The equations of motion reduce to
\begin{eqnarray}
\Lambda_0&=&\Lambda +\fft{(n-4) (n\alpha + \beta)}{n-2}
\Lambda^2\,.\label{vaccon}
\end{eqnarray}
Thus except in four dimensions, there are two disconnect AdS vacua
with different cosmological constants $\Lambda$. (In four
dimensions, there is only one AdS vacuum whose cosmological constant
is the same as the bared $\Lambda_0$.) It can be expected that the
critical condition depends on the vacuum selected. Had we included
the Gauss-Bonnet term in the Lagrangian, the general Einstein
metrics could not be the solution, although the theory would still
admit AdS vacua, with specified cosmological constants determined by
the equations of motion.

       The most general ansatz for a spherically symmetric static
solution is given by
\begin{equation}
ds^2 = -f(r) dt^2 + \fft{dr^2}{g(r)} + r^2
d\Omega_{d}^2\,,\label{bhans}
\end{equation}
where $d=n-2$. For the metric to be Einstein, we must have $f=g$.
The most general solution is the Schwarzschild-Tangherlini black
hole in the AdS background,
\begin{equation}
f=g=1 - \fft{\Lambda}{n-1} r^2 - \fft{2m}{r^{d-1}}\,.\label{fgsol}
\end{equation}
The mass of the black hole can be calculated by using the
Deser-Tekin method \cite{dtmass},
\begin{equation}
{\cal M} =\Big(1 + 2\Lambda (n\alpha + \beta)\Big) {\cal
M}_0\,,\label{bhmass}
\end{equation}
where ${\cal M}_0$ is the mass of the same black hole when it is a
solution of the Einstein theory without higher derivative terms. It
is rather puzzling that the black hole mass, which one expects to be
solely determined by the geometry of the spacetime, would depend on
the theories which it is embedded in.  It is worth remarking that an
identical result of the mass can be also obtained from the AMD
conformal mass formalism \cite{ok,yp}.  Using the Wald formula for
entropy \cite{wald}, we have
\begin{equation}
S=\Big(1 + 2\Lambda (n\alpha + \beta)\Big) S_0\,,\label{bhentropy}
\end{equation}
where $S_0$ is the Bekenstein-Hawking entropy.  If the temperature
of the black hole is unmodified, the first law of black hole
thermodynamics holds in these extended gravities.

    In order to study the stability of the above black hole
solution, we give the general formalism of the linearized analysis
around the background of the Einstein metric (\ref{einstein}) with
the cosmological constant specified in (\ref{vaccon}). Varying the
metric as $g_{\mu\nu}\rightarrow g_{\mu\nu} + h_{\mu\nu}$, and so
$\delta g_{\mu\nu}=h_{\mu\nu}$, we find to first order in variations
that
\begin{eqnarray}
\delta(\cG_{\mu\nu} + \alpha E_{\mu\nu}^{1} + \beta E_{\mu\nu}^{2})
&= &[1+2\Lambda(n\alpha + \beta)]\, \widetilde\cG^L_{\mu\nu} -
\beta\,(\Delta-2\Lambda) \widetilde\cG^L_{\mu\nu}\cr 
&&+ (2\alpha+\beta)\, [-\nabla_\mu\nabla_\nu + g_{\mu\nu}\, \square
+ \Lambda\, g_{\mu\nu}]R^L\,,\label{deom} \end{eqnarray}
where $\widetilde \cG^L_{\mu\nu}$ and $R^L$ are linearized
variations of $\widetilde\cG_{\mu\nu}$ and $R$:
\begin{eqnarray}
\widetilde\cG_{\mu\nu}^L &=& R^L_{\mu\nu} -\ft12 R^L\, g_{\mu\nu}
-\Lambda\, h_{\mu\nu}
\,,\label{GL}\\
R^L_{\mu\nu} &=& \nabla^\lambda\nabla_{(\mu} h_{\nu)\,\lambda}
-\ft12\square h_{\mu\nu} -\ft12 \nabla_\mu\nabla_\nu h\,,
\label{RicL}\\
R^L&=& \nabla^\mu\nabla^\nu h_{\mu\nu} -\square h -\Lambda
h\,.\label{RL} \end{eqnarray}
We have also defined $R^L_{\mu\nu}$ as the linearization of
$R_{\mu\nu}$, and introduced $h=g^{\mu\nu} h_{\mu\nu}$. Note that
$\Delta$ is the Lichnerowicz operator, whose action on a two index
symmetric tensor $T_{\mu\nu}$ in the Einstein space is given by
\begin{eqnarray}
\Delta T_{\mu\nu} &\equiv& -\square T_{\mu\nu} -
2R^{\rho}{}_{\mu\sigma\nu} T_{\rho}{}^{\sigma} + 2R_{(\mu}{}^\rho
T_{\nu)\rho}\cr &=& -\square T_{\mu\nu} - 2R^{\rho}{}_{\mu\sigma\nu}
T_{\rho}{}^{\sigma} + 2\Lambda T_{\mu\nu}\,.
\end{eqnarray}
It is easy to verify that
\begin{equation}
g^{\mu\nu} \widetilde\cG_{\mu\nu}^L = -\ft12(n-2) R^L\,.
\end{equation}
It follows that the trace part of  (\ref{deom}) is given by
\begin{equation}
\Big[\ft12\Big(4(n-1)\alpha + n\beta\Big) \Box - (n-4)
(n\alpha+\beta) \Lambda -\ft12(n-2)\Big] R^{L}=0\,.\label{boxRL}
\end{equation}
In gravity with a cosmological constant, it is convenient to use
general coordinate invariance to impose the gauge condition
\cite{strom}
\begin{equation}
\nabla^\mu h_{\mu\nu}= \nabla_\nu h\,.\label{gauge}
\end{equation}
It follows from (\ref{RL}) that
\begin{equation}
R^L= -\Lambda\, h\,.
\end{equation}
Substituting this into (\ref{boxRL}), we see that the spin-0 mode
$h$ is massive with a mass
\begin{equation}
m_0^2 \sim \fft{1}{4(n-1)\alpha + n\beta}\,.
\end{equation}
We consider a special case when $m_0^2$ becomes infinity and hence
the spin-0 mode decouples from the spectrum.  This corresponds to
\begin{equation}
4(n-1)\alpha + n\beta =0\,,\label{alphabeta}
\end{equation}
for which the equation (\ref{boxRL}) implies that $h=0$. It is this
case, where (\ref{alphabeta}) holds, that we shall focus on in our
subsequent discussion.  In $n=3$, we have $8\alpha+3\beta=0$, and
the resulting theory is the new massive gravity in three dimensions
\cite{bhrt1}.  In $n=4$, we have $3\alpha+\beta=0$, and the
combination of the curvature squared terms form the Weyl-square
term, up to certain total derivatives \cite{lpcritical}.  In all
dimensions, the parameter choice (\ref{alphabeta}) enables one to
write the Lagrangian, with an auxiliary field, in the Pauli-Fierz
form \cite{bhrt}.

Having imposed (\ref{alphabeta}), and hence determined that $h=0$,
we are left with a result that the variation of field equations is
\begin{equation}
0=\delta(\cG_{\mu\nu} + \alpha E_{\mu\nu}^{1} + \beta
E_{\mu\nu}^{2}) =  \fft{2(n-1) \alpha}{n}\,(\Delta -2\Lambda)
(\Delta -2 \Lambda + M^2)\, h_{\mu\nu}\,, \label{hTTeom}
\end{equation}
where $h_{\mu\nu}$ is in the transverse traceless gauge
\begin{equation}
\nabla^\mu\, h_{\mu\nu}=0\,,\qquad g^{\mu\nu}\,
h_{\mu\nu}=0\,,\label{TTgauge}
\end{equation}
and $M^2$ is given by
\begin{equation}
M^2 = \fft{(n-2)^2\Lambda}{2(n-1)} + \fft{n}{4(n-1)\alpha}\,.
\end{equation}

   The most general solution of the fourth-order equation
(\ref{hTTeom}) is a linear combination of a massless graviton,
satisfying
\begin{equation} (\Delta -2\Lambda)\, h^{(m)}_{\mu\nu}=0\,,\label{massless}
\end{equation}
and a massive spin-2 field, satisfying
\begin{equation}
(\Delta -2\Lambda + M^2)\, h^{(M)}_{\mu\nu}=0\,. \label{massive}
\end{equation}
In the AdS background with the cosmological constant $\Lambda$, the
criterion of stability for spin-2 modes in the background requires
that $M^2\ge0$ (see, for example, \cite{kkrep}). Since $\Lambda$ is
negative, we must have
\begin{equation} 0<\alpha \le \fft{n}{4(n-2)(-\Lambda)}\qquad
\longrightarrow \qquad \infty
>M^2\ge 0\,.\label{betarange} \end{equation}
In particular, $\alpha$ must be positive.  This condition ensures
that there is no tachyon instability of the AdS vacuum.

We now examine the stability of the black hole solution given by
(\ref{bhans}) and (\ref{fgsol}). The stability of such a solution in
standard cosmological Einstein gravities in arbitrary dimensions
with a generic Einstein level surface was studied in \cite{gibhar}.
It was argued that the most dangerous mode that can produce
instability is the transverse traceless tensor mode on the level
surface. Since the linearized equation of motion for the massless
graviton (\ref{massless}) is exactly the same as that in Einstein
gravity, and the equation for the massive graviton is modified by
the mass parameter $M$, it is reasonable to expect that the
dangerous modes for the tachyon instability are the same as those
discussed in \cite{gibhar}.  These modes are defined by
\begin{eqnarray}
&&h_{00}=0\,,\qquad h_{0r}=0\,,\qquad h_{0,\alpha}=0\,,\qquad
h_{rr}=0\,,\qquad h_{r\alpha}=0\,,\cr%
&&h_{\alpha\beta}= \tilde h_{\alpha\beta} r^{\fft{4-d}{2}}\Phi(r)
e^{{\rm i} \omega t}\,,
\end{eqnarray}
where $\tilde h_{\alpha\beta}$ is spherical harmonic in
$d\Omega_d^2$, satisfying
\begin{equation}
\tilde \Delta \tilde h_{\alpha\beta} = \lambda \tilde
h_{\alpha\beta}\,.
\end{equation}
The eigenvalues $\lambda$ were obtained in \cite{rubord,hig}. They
are given by
\begin{equation}
\lambda = (\ell + d-1) (\ell + 4)\,,
\end{equation}
with $\ell=0,1,\cdots$. Following the procedure in \cite{gibhar},
the function $\Phi$ satisfies the Schr\"odinger equation
\begin{equation}
-\fft{d^2\Phi}{dr_*^2} + V(r(r_*)) \Phi = \omega^2 \Phi\,.
\end{equation}
where $r_*$ is the Regge-Wheeler type of coordinate, defined by
\begin{equation}
dr_*=\fft{dr}{f}\,.
\end{equation}
The potential in the coordinate $r$ is given by
\begin{equation}
V(r)=\fft{\lambda f}{r^2} + \ft12 (d-4) \fft{f'f}{r} + \ft14
(d^2-10d - 8) \fft{f^2}{r^2} + (M^2-2\Lambda) f\,.
\end{equation}
For the solution to be absent of the tachyon instability, the
potential should ensure that there is no bound state with negative
energy $\omega^2$.  Substituting the black hole solution
(\ref{fgsol}) into the potential, we find that
\begin{equation}
V=\fft{f}{4r^{d+2}}\Big( (4M^2 + d^2 + 2d)r^{d+2} + (4\lambda +
d^2-10d + 8) r^d + d^2 r_+^{d-1} (1 + r_+^2) r\Big)\,,
\end{equation}
where $r_+$ denotes the horizon of the black hole. For $M^2>0$, the
potential is more positive and hence the solution has better
stability. Since we have both $M^2>0$ and $M^2=0$ modes in our case,
the criterion for the absence of the tachyon instability is the same
as the one given in \cite{gibhar}. In particular, for spherical
level surfaces, there is no tachyon instability.  The black hole is
stable even if the sphere is replaced by other Einstein spaces, as
long as the minimum eigenvalue for the tensor harmonic is no less
than the following critical value \cite{gibhar}:
\begin{equation}
\lambda_{\rm crit} \equiv 4 - \fft{(5-d)^2}{4}\,.
\end{equation}
This condition is clearly satisfied by spheres.  Thus there is no
tachyon instability for spherical black holes. The tachyon
instability analysis of black holes with the spherical level surfaces
replaced by other Einstein metrics is the same as that of the usual
gravity studied in \cite{gihapo}.

  For black holes in higher-derivative gravity, even if the tachyon
instability is absent, one still anticipates instabilities
associated with the ghost massive graviton. At the linearized level,
there is no exponentially-growing behavior associated with the ghost
modes; however, their interaction with the normal modes can cause
instability, even at the classical level, by transferring energy
indefinitely between the normal and ghost sectors.

   As we have seen earlier, due to fluctuations of the metric around
the Einstein space background, there are massless and massive spin-2
graviton modes. We now examine the on-shell energy of these modes.
In the case of AdS background, a procedure for doing this has been
first described in \cite{strom} for three-dimensional topologically
massive gravity, based upon the construction of a Hamiltonian for
the graviton field. Generalizations to four and higher dimensions
were given in \cite{lpcritical} and \cite{dllpst}, respectively.
For our black hole backgrounds, the procedure is analogous, although
it should be modified in order to take into account the fact that
the Riemann tensor is no longer trivial. Leaving the parameter
$\alpha$ unrestricted, we may write down the quadratic action whose
variation yields the equations of motion (\ref{hTTeom}):
\begin{eqnarray} I_2 &=& -\fft{1}{2\kappa^2} \, \int\sqrt{-g}\,
d^nx\,h^{\mu\nu}(\delta\cG_{\mu\nu} +
   \delta E_{\mu\nu})\\
&=&-\fft{(n-1) \alpha}{n\kappa^2} \, \int\sqrt{-g}\,
d^nx\,h^{\mu\nu} (\Delta -2\Lambda) (\Delta -2 \Lambda +
M^2)\,h_{\mu\nu} \cr 
&=& -\fft{(n-1) \alpha}{n\kappa^2}\, \int \sqrt{-g}\, d^nx\, \Big[
(\square h^{\mu\nu})(\square h_{\mu\nu}) + M^{2} (\nabla^\lambda
h^{\mu\nu})(\nabla_\lambda h_{\mu\nu}) \cr &&\qquad\qquad\qquad - 4
\Big((\nabla^\lambda h_{\mu\nu})R_{\rho\mu\sigma\nu}(\nabla_\lambda
h^{\rho\sigma}) + h^{\mu\nu}(\nabla^\lambda
R_{\rho\mu\sigma\nu})(\nabla_\lambda h^{\rho\sigma})\Big)\cr 
&&\qquad\qquad\qquad + h^{\mu\nu}(4 M^{2} R_{\rho\mu\sigma\nu}
R_{\delta}{}^{\rho}{}_{\lambda}{}^{\sigma} - 2 M^{2}
R_{\delta\mu\lambda\nu})h^{\delta\lambda}\Big]\,. \end{eqnarray}
Using the method of Ostrogradsky for Lagrangians written in terms of
second, as well as first, time derivatives we define the conjugate
``momenta''
\begin{eqnarray}
\pi^{(1)\mu\nu} &=& \fft{\delta L_2}{\delta \dot h_{\mu\nu}} -
\nabla_0 \Big(\fft{\delta L_2}{\delta(d(\nabla_0
h_{\mu\nu})/dt)}\Big) \nn \\
&=&  -\fft{2(n-1)\alpha}{n\kappa^2}\, \sqrt{-g}\,\Big(  M^{2}
\nabla^0 h^{\mu\nu} - 4 R_{\rho}{}^{\mu}{}_{\sigma}{}^{\nu} \nabla^0
h^{\rho\sigma} - 4 (\nabla^0 R_{\rho}{}^{\mu}{}_{\sigma}{}^{\nu})
h^{\rho\sigma} - \nabla^0 \square h^{\mu\nu}\Big)\,,\nn\\
\pi^{(2)\mu\nu} &=& \fft{\delta L_2}{\delta (d(\nabla_0
h_{\mu\nu})/dt)}=
 -\fft{2(n-1)\alpha}{n\kappa^2}\, \sqrt{-g}\, g^{00}\,
   \square h^{\mu\nu}\,.
\end{eqnarray}
Since the Lagrangian is time-independent, the Hamiltonian is equal
to its time average. It is advantageous to writing it in this way,
because we can then integrate time derivatives by parts.  Thus we
obtain the Hamiltonian
\begin{eqnarray}
H &=& T^{-1}\Big(\int d^n x\, \Big[\pi^{(1)\mu\nu}\, \dot h_{\mu\nu}
+\pi^{(2)\mu\nu}\, \fft{\del(\nabla_0 h_{\mu\nu})}{\del t} \Big] -I_2\Big)\nn\\
&=& -\fft{2(n-1)\alpha}{n\kappa^2T}\, \int \sqrt{-g}\, d^nx\, \Big[
M^{2}(\nabla^0 h^{\mu\nu}) \dot h_{\mu\nu} + 2 (\nabla^0 h^{\mu\nu})
\fft{\partial}{\partial t}\Big( (\Delta - 2\Lambda) h_{\mu\nu}\Big)
\cr 
&&\qquad\qquad\qquad\qquad -4 (\nabla^0
R_\rho{}^\mu{}_{\sigma}{}^{\nu}) h^{\rho\sigma} \dot h_{\mu\nu}
\Big] -\fft1{T}\, I_2\,,\label{H1}
\end{eqnarray}
where all time integrations are taken over the interval $T$.  Not
all tensor components $\nabla^0 R_\rho{}^\mu{}_{\sigma}{}^{\nu}$
vanish for the general static ansatz (\ref{bhans}). However, for the
Einstein metrics where $f=g$, they all vanish.

Evaluating (\ref{H1}) for the massless mode (satisfying
(\ref{massless})) and for the massive mode (satisfying
(\ref{massive})), we obtain the on-shell energies
\bea E_{\rm massless} &=& -\fft{2(n-1)\alpha}{n\kappa^2T} M^{2}
   \int\sqrt{-g}\, d^nx\, (\nabla^0 h^{\mu\nu}_{(m)})\,
   \dot h^{(m)}_{\mu\nu}\,,
\label{masslessE}\\
E_{\rm massive} &=& \fft{2(n-1)\alpha}{n\kappa^2T} M^{2}
   \int\sqrt{-g}\, d^nx\, (\nabla^0 h^{\mu\nu}_{(M)})\,
   \dot h^{(M)}_{\mu\nu}\,.
\label{massiveE} \eea
Thus for generic $M^2\ne 0$, ghost excitation is unavoidable.  In
order for the theory to be free of ghosts, we need to choose the
parameter $\alpha$ such that $M^2=0$.  This is precisely the
critical point where the massive spin-2 field also becomes massless.
It is of interest to note that the critical condition for black
holes is the same as that for the vacuum. Note also that the black
hole mass (\ref{bhmass}) and entropy (\ref{bhentropy}) are both zero
at this critical point.  The vanishing of black hole masses at
critical points was first observed in three dimensions
\cite{lupang}, and subsequently in four \cite{lpcritical} and
general dimensions \cite{dllpst}.  The only modes left to discuss
are the log modes, which can be obtained from the limit
\begin{equation}
h^{\log}_{\mu\nu} = \lim_{M\rightarrow 0} \fft{h_{\mu\nu}^{(M)} -
h_{\mu\nu}^{(m)}}{M^2}\,.\label{logdef}
\end{equation}
It is clear that the log modes satisfy
\begin{equation}
(\Delta - 2\Lambda) h_{\mu\nu}^{\log} = -
h_{\mu\nu}^{(m)}\,.\label{logeom}
\end{equation}
Thus we see that $(\Delta - 2\Lambda) h_{\mu\nu}^{\log}\ne 0$ but
$(\Delta - 2\Lambda)^2 h_{\mu\nu}^{\log}=0$. Substituting
(\ref{logeom}) into (\ref{H1}), the on-shell energy for the log
modes can be expressed as
\begin{equation}
E_{\log} = \fft{4(n-1)\alpha}{n\kappa^2 T} \int \sqrt{-g} d^n x\,
(\nabla^0 h^{\mu\nu}_{\log})\, \dot h^{(m)}_{\mu\nu}\,.\label{elog}
\end{equation}
We do not expect to find exact analytical solution for these log
modes in the black hole background.  The log modes in AdS backgrounds
were constructed \cite{af,ggst,bhrt}. Their on-shell energies are
evaluated in the appendix for $D=3$ and $D=4$.

The most general linearized solution at the critical point is thus
given by
\begin{equation}
h_{\mu\nu} = c_1 h^{\log}_{\mu\nu} + c_2
h^{(m)}_{\mu\nu}\,,\label{genmodes}
\end{equation}
where $c_1$ and $c_2$ are arbitrary constants. We find that the
energy is given by
\begin{eqnarray}
E(c_1,c_2) &=& c_1^2 E_{\log} + 2 c_1 c_2 E_{\rm cross}\cr 
&=& \Big(c_1 + \fft{E_{\rm cross}}{E_{\log}}\,c_2\Big)^2 E_{\log} -
c_2^2\, \fft{E_{\rm cross}^2}{E_{\log}}\,,\label{genenergy}
\end{eqnarray}
where
\begin{equation}
E_{\rm cross} \equiv \fft{2(n-1)\alpha}{n\,\kappa^2 T}\int \sqrt{-g}
d^n x\, (\nabla^0 h^{\mu\nu}_{(m)})\, \dot
h^{(m)}_{\mu\nu}\,.\label{ecross}
\end{equation}
Note that the cross term $E_{\rm cross}$ vanishes only for the
``Proca'' log modes where $h_{\mu\nu}^{(m)}$ is pure gauge, {\it
i.e.} $h_{\mu\nu}=\nabla_{\mu}\xi_\nu + \nabla_{\nu} \xi_\mu$. Thus
we see that certain mixed states of the massless and log modes have
negative energy. This conclusion, which also applies to the AdS
vacua, is equivalent to the analysis of the scalar product of these
modes in the AdS background \cite{porrati}.

To summarize, we have considered extended cosmological gravities in
diverse dimensions with Ricci scalar and tensor squared terms. These
theories were known previously to have critical points where the
ghost massive graviton disappears.  We showed that these theories
admit solutions of Einstein metrics, including both the AdS vacua
and Schwarzschild-Tangherlini AdS black holes. We performed
linearized analysis around the black hole background and showed that
the spectrum consists of the usual spin-2 massless and ghost massive
gravitons. We first argued that there is no exponentially-growing
tachyon mode in the black holes. Then we examined critical points,
where the ghost massive graviton is replaced by the log modes, which
have non-vanishing energy, with the sign choice depending on the
overall sign of the action. We showed that there always exist
certain mixing of the massless and log modes that have negative
energy.  The stability of the black holes thus requires that the log
modes to be truncated out by the boundary condition. (Note that
since we have performed the linearized analysis around the generic
Einstein metrics, we expect our results apply to the stability of
the general Kerr-AdS solutions \cite{carter,hht,glpp} as well.  In
fact, it can be checked that the quantities $\nabla^0R_\rho {}^\mu
{}_\sigma{}^\nu$ in (\ref{H1}) vanish for the Kerr-AdS solutions.)
This conclusion applies to the stability of the AdS vacua as well.
This is different from the critical topologically massive gravity
where log modes can be kept in one Virasoro sector, but not the
other \cite{mss}.

Although the boundary condition that removes the log modes may render
critical gravities to appear trivial with null modes only, these
remaining massless modes are the standard graviton in the usual
Einstein theory of gravity.  This is different from the case in
three dimensions where the massless modes are indeed pure gauge.
Analogously, the non-trivial Schwarzschild-Tangherlini AdS black
holes acquire zero mass and entropy in critical gravities. Further
investigation of these unusual properties may shed light on the dual
log CFT at higher temperature.

\section*{Acknowledgment}

We are grateful to Sera Cremonini, Yi-Hong Gao, Chris Pope and
Zhao-Long Wang for useful discussions. H.Liu and ML are supported in
part by the National Science Foundation of China (10875103),
National Basic Research Program of China (2010CB833000), and
Zhejiang University Group Funding (2009QNA3015). H.L.~would like to
thank the organizer and the participants of the advanced workshop
``Dark Energy and Fundamental Theory'', supported by the Special
Fund for Theoretical Physics from the National Natural Foundation of
China for useful discussions.

\appendix

\section{Explicit energy of spin-2 modes in $D=3,4$ AdS vacua}

     The massive and massless graviton modes in the AdS$_3$ background
\begin{equation}
ds_3^2= -\cosh^2\rho\, dt^2 + \sinh^2\rho d\phi^2 + d\rho^2
\end{equation}
were obtained in \cite{strom}.  The ansatz is given by
\begin{eqnarray}
\psi_{\mu\nu}(h,\bar h) &=& e^{-{\rm i}(h + \bar h)t - {\rm i} (h -
\bar h) \phi} \fft{\sinh^2\rho}{(\cosh\rho)^{h + \bar h}}
\begin{pmatrix} 1 & \ft12 (h-\bar h) & \fft{2 \rm
i}{\sinh(2\rho)}\cr 
\ft12(h-\bar h) & 1 & \fft{{\rm i} (h-\bar h)}{\sinh(2\rho)} \cr 
\fft{2 \rm i}{\sinh(2\rho)} & \fft{{\rm i} (h-\bar h)}{\sinh(2\rho)}
& - \fft{4}{\sinh^2(2\rho)}
\end{pmatrix}\,,
\end{eqnarray}
where $h$ and $\bar h$ are constants, $\psi_{\mu\nu}$ is traceless.
The transversality condition is satisfied provided that
\begin{equation}
h-\bar h=\pm 2\,.\label{hbh}
\end{equation}
Let us first consider the upper sign choice, {\it i.e.} $\bar h =
h-2$. Then $\psi_{\mu\nu}$ is a massive graviton satisfying
\begin{equation}
(\Box + 2 - M^2) \psi_{\mu\nu}=0\,,
\end{equation}
where
\begin{equation}
M^2=4(h-1)(h-2)\,.
\end{equation}
For $M^2>0$, we have $h\le 1$ or $h\ge 2$. The $h\le 1$ branch is
ruled out by the boundary condition.  Thus we must have $h\ge 2$,
with $h=2$ corresponding the massless graviton.  For generic $h$, we
have
\begin{eqnarray}
&&\int\sqrt{-g}\, d^3x\, (\nabla^0 h^{\mu\nu}_{(m)})\,
   \dot h^{(m)}_{\mu\nu}\cr 
   &&\qquad\qquad= 2\pi T \int_0^\infty d\rho
\fft{4 (h-1) (1 - 2 h + \cosh(2\rho))
\sinh\rho}{(\cosh\rho)^{4h+1}}\,.
\end{eqnarray}
For $h\ge 2$, the integrand is negative.  In particular, the
integral is $-4\pi T/3$ when $h=2$.  It follows from
(\ref{masslessE}) and (\ref{massiveE}) that massless modes have
positive energy whilst massive modes have negative energy.  In
calculating the energy, we have chosen the real part of
$\psi_{\mu\nu}$ as the graviton field.  The result is identical if
we have chosen the imaginary part.

    The log modes, defined by (\ref{logdef}), are given by
\begin{equation}
\psi_{\mu\nu}^{\log} =- \ft12 \Big({\rm i}\, t +
\log(\cosh\rho)\Big) \psi_{\mu\nu}(2,0)\,.
\end{equation}
Thus we have
\begin{eqnarray}
&&\int \sqrt{-g} d^3 x\, (\nabla^0 h^{\mu\nu}_{\log})\, \dot
h^{(m)}_{\mu\nu}\cr 
&&\qquad\qquad =-4\pi T \int_0^\infty d\rho \tanh\rho\, {\rm
sech}^8\rho \Big(1 + (\cosh(2\rho) -3) \log(\cosh\rho)\Big)
\end{eqnarray}
which yields $-17\pi T/36$. It follows from (\ref{elog}) that the
on-shell energy of log modes is negative for the standard sign
choice of the Einstein-Hilbert action, but becomes positive if we
reverse the sign of the action.  The conclusion is the same for the
lower sign choice in (\ref{hbh}).

The situation is somewhat different in four dimensions. The log modes was
shown to have positive energy in \cite{lpcritical}.  Here we present
some detail. The massive and massless graviton modes in the AdS$_4$
background
\begin{equation}
ds_3^2= -\cosh^2\rho\, dt^2 + \sinh^2\rho (d\theta^2 + \sin^2\theta
d\phi^2) + d\rho^2
\end{equation}
were obtained in \cite{bhrt}. They are given by
\begin{eqnarray} &&\psi_{\tau\tau} =
-\psi_{\tau\phi}=\psi_{\phi\phi}=e^{-3{\rm i} \tau + 2{\rm i}\phi}
\sin^2\theta \sinh^{-\fft12} (2\rho)
\tanh^{\fft52}\rho\,,\cr 
&&\psi_{\tau\rho} = - \psi_{\rho\phi} = -{\rm i} {\rm csch}\rho\,
{\rm sech}\rho\, \psi_{\tau\tau}\,,\cr 
&&\psi_{\tau\theta}=-\psi_{\theta\phi} = {\rm i} \cot\theta\,
\psi_{\tau\tau}\,,\qquad \psi_{\rho\rho} = - {\rm csch}^2(2\rho)\,
\psi_{\tau\tau}\,,\cr 
&&\psi_{\rho\theta} = -2\cot\theta\, {\rm csch}(2\rho)\,
\psi_{\tau\tau}\,,\qquad \psi_{\theta\theta} = -
\cot^2(\theta)\psi_{\tau\tau}\,,
\end{eqnarray}
They are traceless and transversal and satisfy
\begin{equation}
(\Box + 2 - M^2) \psi_{\mu\nu}=0\,, \qquad {\rm with}\qquad M^2=
E_0(E_0 -3)\,.
\end{equation}
To obtain the on-shell energy, we first need to evaluate the
integrand
\begin{eqnarray}
&&\int\sqrt{-g}\, d^4x\, (\nabla^0 h^{\mu\nu}_{(M)})\,
   \dot h^{(M)}_{\mu\nu}\cr 
   &&\qquad= 2\pi T \int_0^\infty d\rho
\Big[-\ft2{15} E_0 \Big(68 + 89 E_0 + 4(7E_0 - 16) \cosh(2\rho)\cr
&&\qquad\qquad + (3E_0-4)\cosh(4\rho)\Big){\rm sech}^3\rho\,
(\sinh(2\rho))^{-E_0} (\tanh\rho)^{E_0+2}\Big]\,.
\end{eqnarray}
For $E_0\ge 3$, the integrand is negative.  In particular, when the
above quantity is $-9\pi^2 T/16$ when $E_0=3$. Thus the energy of
the massless graviton is positive and that of the massive graviton
is negative for non-critical gravity.

   The log modes, defined by (\ref{logdef}), are given by
\begin{equation}
\psi_{\mu\nu}^{\log} = -\ft16 (2{\rm i}\, t -\log(\sinh(2\rho)) +
\log (\tanh\rho) )\psi_{\mu\nu} (E_0=3)\,.
\end{equation}
Thus we have
\begin{eqnarray}
&&\int\sqrt{-g}\, d^4x\, (\nabla^0 h^{\mu\nu}_{\log})\,
   \dot h^{(m)}_{\mu\nu}\cr 
   &&\qquad= 2\pi T \int_0^\infty d\rho \,P(\rho)\,,
\end{eqnarray}
where
\begin{eqnarray}
P(\rho) &=&\ft{1}{120} \tanh^2\rho\, {\rm sech}^9\rho \Big(-2 (89 +
28 \cosh(2 \rho) + 3 \cosh(4 \rho)) \cr 
&&\qquad 5 (67 + 4 \cosh(2 \rho) + \cosh(4 \rho)) (\log(\sinh(2
\rho)) - \log(\tanh(\rho)))\Big)\,.
\end{eqnarray}
This function is non-negative for $\rho \in [0, \infty)$ and
$\int_0^\infty P d\rho \sim 0.0386968$. It follows from (\ref{elog})
that the on-shell energy of the log modes is positive for the
standard sign of the Einstein-Hilbert term.

      However, in general the most general solution is a linear
combination of the massless and log modes, as in (\ref{genmodes}).
The corresponding energy is of the form (\ref{genenergy}). Thus
unless we consistently truncate out the log modes, there are always
combinations that have negative energies.

\end{document}